\documentclass[aps,twocolumn,prl,showpacs,showkeys,preprintnumbers,superscriptaddress,nobibnotes,
floatfix,longbibliography,notitlepage,
]{revtex4-1}
\pdfoutput=1

\usepackage{amsmath}
\usepackage{amsfonts}
\usepackage{amssymb}
\usepackage{mathrsfs}
\usepackage{cancel}
\usepackage{accents}
\usepackage{mciteplus,slashed}
\usepackage{amssymb,cancel,amsmath,relsize}
\usepackage{mathrsfs} 
\usepackage{dcolumn}
\usepackage{bm}
\usepackage[caption=false]{subfig} 
\usepackage{appendix}
\usepackage{physics}
\usepackage{feynmp-auto}
\usepackage[T1]{fontenc}	
\usepackage{csvsimple}
\usepackage{hyperref}
\usepackage[capitalise]{cleveref}
\usepackage{booktabs}
\usepackage{graphicx}
\usepackage{mathrsfs}
\usepackage[utf8]{inputenc}
\usepackage[T1]{fontenc}
\usepackage[dvipsnames]{xcolor}
\hypersetup{
    colorlinks,
    linkcolor={red!50!black},
    citecolor={blue!50!black},
    urlcolor={blue!80!black}
}
\usepackage[normalem]{ulem}
\usepackage{cleveref}

\renewcommand{\section}[1]{{\noindent \bf{#1.}---}}

\newcommand{\Ztwo}{\ensuremath{\mathcal{Z}_2}}
\newcommand{\ZDM}{\ensuremath{\mathcal{Z}_2^{\rm (DM)}}}
\newcommand{\ZDW}{\ensuremath{\mathcal{Z}_2^{\rm (DW)}}}

\graphicspath{{Figures/}}

\begin{document}

\title{Quantum Gravity Effects on Dark Matter and Gravitational Waves}

\author{Stephen F. King}
\affiliation{School of Physics and Astronomy, University of Southampton, Southampton SO17 1BJ, United Kingdom}

\author{Rishav Roshan}
\affiliation{School of Physics and Astronomy, University of Southampton, Southampton SO17 1BJ, United Kingdom}

\author{Xin Wang}
\affiliation{School of Physics and Astronomy, University of Southampton, Southampton SO17 1BJ, United Kingdom}

\author{Graham White}
\affiliation{School of Physics and Astronomy, University of Southampton, Southampton SO17 1BJ, United Kingdom}

\author{Masahito Yamazaki}
\affiliation{Kavli IPMU (WPI), UTIAS, The University of Tokyo, Kashiwa, Chiba 277-8583, Japan}
\affiliation{Center for Data-Driven Discovery, Kavli IPMU (WPI), UTIAS, The University of Tokyo, Kashiwa, Chiba 277-8583, Japan}
\affiliation{Trans-Scale Quantum Science Institute, The University of Tokyo, Tokyo 113-0033, Japan}

\begin{abstract}
We explore how quantum gravity effects, manifested through the breaking of discrete symmetry responsible for both Dark Matter and Domain Walls, can have observational effects through CMB observations and gravitational waves. To illustrate the idea we consider a simple model with two scalar fields and two $\Ztwo$ symmetries, one being responsible for Dark Matter stability, and the other spontaneously broken and responsible for Domain Walls, where both symmetries are assumed to be explicitly broken by quantum gravity effects. We show the recent gravitational wave spectrum observed by several pulsar timing array projects can help constrain such effects.
\end{abstract}

\maketitle

\section{Introduction}Global symmetries are ubiquitous in Nature, being already present in the Standard Model (SM) of particle physics such as the baryon and lepton numbers. Discrete global symmetries often play a role in many theories beyond the SM, such as Dark Matter (DM) and Neutrino Mass models. Unlike gauge symmetries (this includes gauge discrete symmetries, for example, those that emerge from the Higgsing of a gauged $U(1)$ symmetry), conventional wisdom tells us that such global symmetries should be broken \cite{Banks:1988yz,Banks:2010zn,Harlow:2018tng} in theories of quantum gravity (QG), e.g.\ by wormholes \cite{Giddings:1987cg}. Such ideas fit nicely into recent developments on swampland conjectures \cite{Vafa:2005ui,Ooguri:2006in}, which classify low energy effective field theories (EFTs) according to their compatibility with QG. Although QG is expected to break all global symmetries, the strength of the breaking is not a priori specified. The breaking may be associated with operators of any mass dimension greater than four. The dimensional scale associated with such operators may be equal to the Planck scale $M^{}_{\rm Pl}$~\cite{Addazi:2021xuf, Rai:1992xw,  Mambrini:2015sia}, while the operators may be suppressed by non-perturbative effects leading to an effective breaking scale many orders of magnitude higher.

\begin{figure}[t!]
    \centering
    \includegraphics[width=0.9\linewidth]{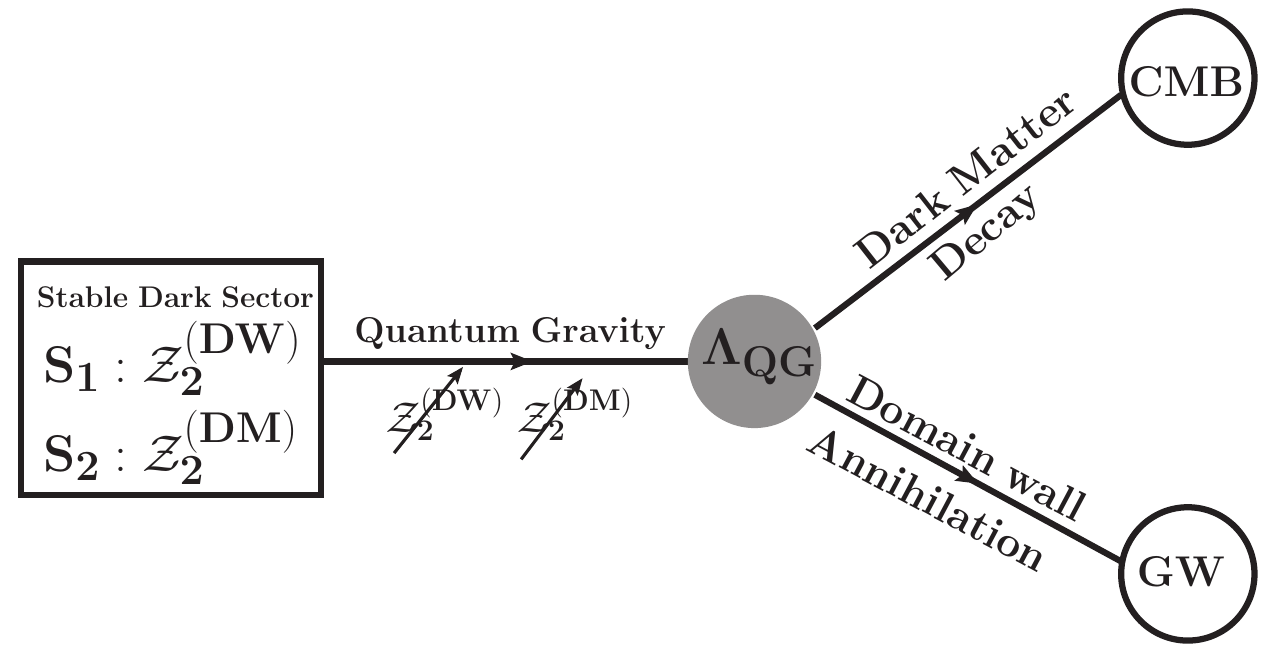}  
    \caption{Schematic of how indirect detection and gravitational wave observatories can provide independent witnesses of the scale of QG which we assume to be approximately common.}
    \label{fig:schematic}
\end{figure}

In this Letter, we explore how QG effects, manifested through the breaking of discrete symmetry responsible for both DM and Domain Walls (DWs), can have observational effects through CMB observations and gravitational waves (GWs). We especially show that QG motivates the existence of very small bias terms which are often assumed to exist to allow DWs to annihilate, thus preventing them from dominating the energy budget of the Universe. To illustrate the idea we consider a simple model with two singlet scalar fields and two $\Ztwo$ symmetries, one being responsible for DM stability, and the other spontaneously broken and responsible for DWs, where both symmetries are assumed to be explicitly broken by QG effects by operators at the same mass dimension and with the same effective Planck scale.
We shall show that this hypothesis leads to observable GW signatures from the DWs annihilation, which are correlated with the decaying DM signatures constrained by CMB observations. The simple set-up described above is depicted in \cref{fig:schematic}.

Recently, several pulsar timing array (PTA) projects reported the discovery of a stochastic gravitational wave background (SGWB), in particular, the North American Nanohertz Observatory for Gravitational Waves (NANOGrav)~\cite{NANOGrav:2023gor, NANOGrav:2023hvm}, the European PTA~\cite{Antoniadis:2023ott, Antoniadis:2023zhi}, the Parkes PTA~\cite{Reardon:2023gzh} and the Chinese PTA~\cite{Xu:2023wog}.
This could be due to the merging of supermassive black hole binaries~\cite{Sesana:2004sp, Burke-Spolaor:2018bvk}, or it may have a cosmological origin or a combination of effects. 
For example, the cosmological origin of SGWB could be due to first-order phase transitions~\cite{Winicour1973, Hogan:1986qda, Athron:2023xlk, Caprini:2010xv, NANOGrav:2021flc, Xue:2021gyq, DiBari:2021dri, Madge:2023cak}, cosmic strings~\cite{Siemens:2006yp, Ellis:2020ena, King:2020hyd, Buchmuller:2020lbh, Blasi:2020mfx, Bian:2022tju, Fu:2023nrn}, or DWs annihilation~\cite{Ferreira:2022zzo, An:2023idh, Dunsky:2021tih, Sakharov:2021dim}, where the latter is of particular interest here. Indeed several papers have appeared which discuss these possibilities~\cite{King:2023cgv, Han:2023olf, Guo:2023hyp, Kitajima:2023cek, Bai:2023cqj, Kitajima:2023cek, Vagnozzi:2023lwo, Ellis:2023tsl, Franciolini:2023pbf, Athron:2023mer, Kitajima:2023vre, Lazarides:2023ksx, Yang:2023qlf, Addazi:2023jvg, Broadhurst:2023tus, Cai:2023dls, Inomata:2023zup, Depta:2023qst, Depta:2023qst, Eichhorn:2023gat, Huang:2023chx, Gouttenoire:2023ftk,Blasi:2023sej,Liu:2023pau,Ahmadvand:2023lpp,Zhang:2023nrs,Jin:2023wri,Antusch:2023zjk,Salvio:2023ynn,Gouttenoire:2023bqy,Du:2023qvj,Li:2023tdx,DiBari:2023upq,Unal:2023srk,Ghosh:2023aum, Jiang:2023gfe, Zhu:2023lbf, An:2023jxf,Borah:2023sbc,Barman:2023fad, Babichev:2023pbf, Ben-Dayan:2023lwd, Liu:2023ymk, Niu:2023bsr, Li:2023bxy,Ge:2023rce,Wang:2023ost}.
One of the goals of the present paper is to investigate the implications of the PTA results on the framework of interest here.

\vspace{0.5 em} 
\section{Discrete Global Symmetry Breakings in QG}In particle physics model building,
it is often useful to invoke discrete global symmetries in 
EFTs. 
It has long been believed~\cite{Banks:1988yz,Banks:2010zn,Harlow:2018tng}, however, that 
there exists no exact (continuous or discrete) global symmetry in QG theories~\footnote{This statement can be violated in QG theories involving ensemble averages, see, e.g.\ refs.~\cite{Antinucci:2023uzq,Ashwinkumar:2023jtz} for recent discussions.}.
In other words, any global symmetry of a given EFT is at best an approximate symmetry emergent in the IR \cite{Witten:2017hdv}, and should be broken by a higher-dimensional operator
\begin{align}
    \mathcal{L}_{\cancel{\Ztwo}}
    = \frac{1}{\Lambda_{\rm QG }} \mathcal{O}_5 \; ,
\label{eq:breaking}
\end{align}
where we consider the leading dimension-five operator in four spacetime dimensions.

One might naively expect that $\Lambda_{\rm QG} \sim \mathcal{O}(M_{\rm Pl})$, since this is a QG effect. 
A global symmetry, however, can be broken by non-perturbative instanton effects (e.g.\ D-brane instanton in string theory \cite{Blumenhagen:2006xt,Florea:2006si,Blumenhagen:2009qh} or gravitational instanton \cite{Giddings:1987cg,Lee:1988ge,Abbott:1989jw,Coleman:1989zu}). The operator in eq.~\eqref{eq:breaking} is then suppressed by a factor $e^{-\mathcal{S}}$, where the dimensionless parameter $\mathcal{S}$ represents the size of the action of the nonperturbative instanton~\footnote{Such a small parameter can also be generated by warped throats  \cite{Randall:1998uk} and conformal sequestering \cite{Luty:2000ec}. It is, however, often non-trivial to realize these scenarios in string theory, see, e.g.\ \cite{Anisimov:2001zz,Anisimov:2002az,Kachru:2006em} for discussions on sequestering from warped throats in string theory.}. In this case, the scale $\Lambda_{\rm QG}$ 
should be estimated as
\begin{align}
    \Lambda_{\rm QG} \sim M_{\rm Pl}\, e^{\mathcal{S}} \gg M_{\rm Pl} \; .
\end{align} 
In principle, one can achieve an effective trans-Planckian scale generated by fields with sub-Planckian mass if there is a very small coupling in the theory. However, the exponential enhancement due to the instanton means we are typically many orders of magnitude above the Planck scale. This implies any resulting phenomenology is almost certainly due to QG effects.
In this Letter, we consider a scenario where a few different 
$\Ztwo$-global symmetries are broken by higher dimensional operators associated with the {\it same} energy scale $\Lambda_{\rm QG}$.
In general, there is no guarantee that different global mechanisms for global symmetry breaking are associated with the same energy scale.
There are, however, important motivations for this assumption, and our discussion can be regarded as a minimal example representing the spirit of more general constraints.

The first motivation comes from the fact that the number of tunable parameters in EFT is finite in string theory---both the number of Calabi-Yau geometries and the choices of fluxes therein are believed to be finite \cite{Acharya:2006zw}, and this leads to infinite constraints on the higher-dimensional operators in the EFT (see, e.g.\ ref.~\cite{Heckman:2019bzm} for a recent discussion). Our assumption is a minimal incarnation of this constraint.

The second motivation comes from a general consideration of the $\Ztwo$-symmetry breaking.
While it has been believed that there are no exact global symmetries in QG, this constraint is not useful unless one formulates a quantitative statement concerning the size of global symmetry breaking. Suppose that we consider a class of string theory compactifications where 
the energy scale $\Lambda_{\rm QG}$ satisfies the inequality
$\Lambda_{\rm QG} \lesssim \Lambda_{\rm max}$.
If such a bound exists, the choice $\Lambda_{\rm QG} \sim \Lambda_{\rm max}$
gives the best attainable quality of global symmetry in the class of string theory compactifications, 
and a model builder is well motivated to choose this value for all higher-dimensional operators whose sizes are severely constrained by experiments~\footnote{More generally, we can formulate the following conjecture (which we call the Global Symmetry Breaking Swampland Conjecture): Let us consider an EFT with a UV completion in theories of QG. Any $\Ztwo$ global symmetry should be broken by a higher-dimensional operator of the form in eq.~\eqref{eq:breaking}, where the energy scale $\Lambda_{\rm QG}$ satisfies the inequality $\Lambda_{\rm QG} \lesssim \Lambda_{\rm max}$ with $\Lambda_{\rm max}$ being a universal scale independent of the choice of the EFT. 
In the main text, we formulated this conjecture for a specific class of string theory compactifications,
which we expect will lead to a smaller value of $\Lambda_{\rm max}$ than that for general EFTs. We can formulate the conjecture for any global symmetry, either continuous or discrete, Abelian or non-Abelian.}.

The consequences of our scenarios depend heavily on the values of $\Lambda_{\rm QG}$. In general, it is often believed the scale of a global symmetry breaking can be much higher than the Planck scale.
For example, in order for a global $U(1)$ Peccei-Quinn (PQ) symmetry  to solve the strong CP problem \cite{Peccei:1977hh,
Peccei:1977ur}, the size of the instanton action $\mathcal{S}$ should 
satisfy $\mathcal{S}\gtrsim 190$, resulting in an extremely high energy scale $\Lambda_{\rm QG}\sim 10^{100}~\textrm{GeV}$ (see refs.~\cite{Kamionkowski:1992mf,Holman:1992us,Barr:1992qq} for early references on the axion quality problem)~\footnote{Estimates on the size of PQ symmetry-breaking operators due to wormholes are also different depending on the choice of theories (see, e.g.\ ref.~\cite{Alvey:2020nyh} and references therein)}. It is, however, non-trivial to have such a high value of $\mathcal{S}$ \cite{Svrcek:2006yi} as there can be many non-perturbative QG effects that can violate a global symmetry and one would need to suppress all of them. In our case, since we are not pushed into a difficult corner of parameter space by the axion quality problem, we wish to consider what seems to be a more natural range of $\mathcal{S}$, that is lower, with the caveat that the precise value will depend upon the choice of string compactification.


At present we know little about the case of discrete $\Ztwo$-symmetries considered in this Letter. However, general estimates
suggests that the size of the non-perturbative instanton action scales as  
$\mathcal{S}\sim \mathcal{O}(M_{\rm Pl}^2/\Lambda_{\rm UV}^2)$ \cite{Fichet:2019ugl,Daus:2020vtf}, where $\Lambda_{\rm UV}$ is the UV cutoff of the theory.
One can consider a scenario where $\Lambda_{\rm UV} \lesssim M_{\rm Pl}$, which could generate a value of $\mathcal{S}\sim \mathcal{O}(10)$. In the following we consider the energy scale $\Lambda_{\rm QG}\sim (10^{20} \cdots 10^{35})~\textrm{GeV}$, which corresponds to the value $\mathcal{S}\sim (4 \cdots 38)$.
In practice, we can keep $\Lambda_{\rm QG}$ as a free parameter
whose value can be constrained by phenomenological and cosmological considerations.

\vspace{0.5 em} 
\section{Simplified Model for BSM Scenarios}In the rest of this Letter, we consider
a minimalistic model where the Standard Model is extended by two singlet scalar fields $S_1$ and $S_2$, each subject to a $\Ztwo$-global symmetry, which generates the following scalar potential at tree level
\begin{eqnarray}
    V&=& \mu ^2  H^\dagger H+ \lambda (H^\dagger H)^2 +H^\dagger H (\lambda _{hs1}S^2_1 + \lambda_{hs2} S_2^2) \nonumber \\ && 
    + \lambda _{s12} S_1^2 S_2 ^2   + \mu_2^2 S_2^2 + \frac{\lambda_2}{4} S_2^4+\frac{\lambda_1}{4} (S_1^2-v_1^2)^2   \ .
\end{eqnarray}
Here $H$ is the SM Higgs doublet field, and the singlet under the SM gauge group $S_2$ is our DM candidate, protected by an approximate $\Ztwo$-symmetry $\ZDM$.
The field $S_1$ is another scalar singlet under the SM gauge group with an approximate $\Ztwo$-symmetry $\ZDW$, and this field will acquire a vacuum expectation value (vev) $v^{}_1$ in the early Universe, generating DWs in the process. The scalar potential should be bounded from below to make the electroweak vacuum stable, which poses constraints on the scalar couplings ~\cite{Bhattacharya:2019tqq}. Next, we write the mass of $S_2$ as $m_{2}^2=2\mu_2^2+\lambda _{hs1}v_h^2+2\lambda _{s12}v_1^2$ with $v_h=246$ GeV being the vev of $H$, and consider $m^2_{2}$ to be positive throughout to avoid the inclusion of a non-trivial vev of $S_2$. We also assume $\lambda^{}_{hs1}$ to be sufficiently small so that there would not be large mixing between $H$ and $S_1$.

The two $\mathcal{Z}_2$-symmetries are however broken by higher-dimensional operators of the form 
    \begin{eqnarray}
        \Delta V  &=& \frac{1}{\Lambda_{\rm QG}} \sum _{i=1}^2 (\alpha_{1i}  S_i^5 + \alpha_{2i}  S_i^3 H^2 + \alpha_{3i}  S_i H^4 ) \nonumber \\ 
        && + \frac{1}{\Lambda _{\rm QG}} \sum _{j=1}^4 c_{j} S^j _1 S^{5-j}_2\ . 
        \label{eq:bias}
    \end{eqnarray}
As discussed before, we assume a common origin and therefore a common scale for the breaking of all global symmetries, hence we simply take all the dimensionless coefficients in eq.~\eqref{eq:bias} to be of the same order, and we can make them of ${\cal O}(1)$ by redefining $\Lambda_{\rm QG}$. Let us note here that in string theory compactifications one often encounters a large number $\mathcal{O}(100)$ of SM-singlet moduli fields, and this suggests that the scalar DM can be one of such fields. The effect of these fields in cosmology is often discussed in the context of the cosmological moduli problem \cite{Coughlan:1983ci,Goncharov:1984qm,Ellis:1986zt,Banks:1993en,deCarlos:1993wie}: the late-time decay of the moduli spoils the success of the Big-Bang Nucleosynthesis (BBN), and this requires the mass of the moduli fields to be above $\mathcal{O}(10)$ TeV~\footnote{In $\Delta V$ we have absorbed factors of $\sqrt{N}$ from the number of moduli fields into $\Lambda _{\rm QG}$}. While we are considering a simplified model with only two scalar fields, it is reasonable to have the DM abundance dominated by the contribution of a single particle species, and the GW spectrum from DWs annihilation is typically dominated by the contribution of a single scalar field, as we will explain below. Therefore, our simple model captures the qualitative features of a large class of string-inspired models.

It is interesting to point out that, once the electroweak symmetry breaking is triggered, the operator $S_2 H^4/ \Lambda_{\rm QG}$ present in eq.~\eqref{eq:bias} allows $S_2$ to mix with the CP even scalar component of $H$ \footnote{In principle, $S_2$ can mix with $S_1$ while $S_1$ can mix with the CP even scalar of $H$. As these extra mixings will not affect our phenomenology in great detail, we remain agnostic about them in the rest of the analysis. }. We identify one of the physical scalars obtained after mixing as the SM Higgs with $m_h=125$ GeV while the other physical scalar plays the role of the DM with mass $m_\text{DM}$. This scalar mixing can be parameterized as
\begin{equation}
   \sin{\theta}=\frac{v_h^3}{(m_h^2-m_\text{DM}^2)\Lambda^{}_{\rm QG}} \ .
   \label{scalar-mixing}
\end{equation}
As a result of this mixing the DM can decay to all the SM particles. The expression of the DM decay width can be found in, e.g. ref.~\cite{Spira:2016ztx}. The lifetime of such decay is highly constrained from CMB, which will be discussed in the following.

\vspace{0.5 em} 
\section{Indirect detection of QG scale}
The explicit breaking of $\ZDM$-symmetry by higher-dimensional operators originating from QG effects in the present set-up allows the DM to decay into SM particles. These SM particles can further decay to photons, electron-positron and neutrino-antineutrino pairs.
If the DM decay happens during or after the era of recombination, the energy injected can reionize the intergalactic medium and modify the CMB power spectrum. Remarkably the resulting limits on the DM lifetime tend to be much larger than the age of the Universe $\tau_\text{DM}\gtrsim 10^{25}$~s for an ${\cal O}(1)$ branching ratio into electromagnetic final states and high efficiency into ionization  channels~\cite{Slatyer:2016qyl,Planck:2018vyg}. This results in a bound on $\Lambda _{\rm QG}$ well above the Planck scale.

On the other hand, the $e^+e^-$ pairs, if produced during the DM decays originating from the DM-dominated galaxies and clusters, can undergo energy loss via electromagnetic interactions in the interstellar medium and can give rise to radio waves. Such radio signals can then be observed by several radio telescopes. The Square Kilometer Array (SKA) radio telescope is one such example~\cite{Colafrancesco:2015ola}. SKA provides a much better probe for the decaying DM parameter space in comparison to the existing gamma-ray observations. A recent study~\cite{Dutta:2022wuc}, found that the DM decay width $\Gamma_\text{DM}\gtrsim10^{-30}~{\rm s}^{-1}$ is detectable at SKA (assuming 100 hours of observation time). This, in turn, suggests that the QG effects can also be tested by SKA.

\begin{figure*}[t!]
  \centering
  \includegraphics[width=0.32\linewidth]{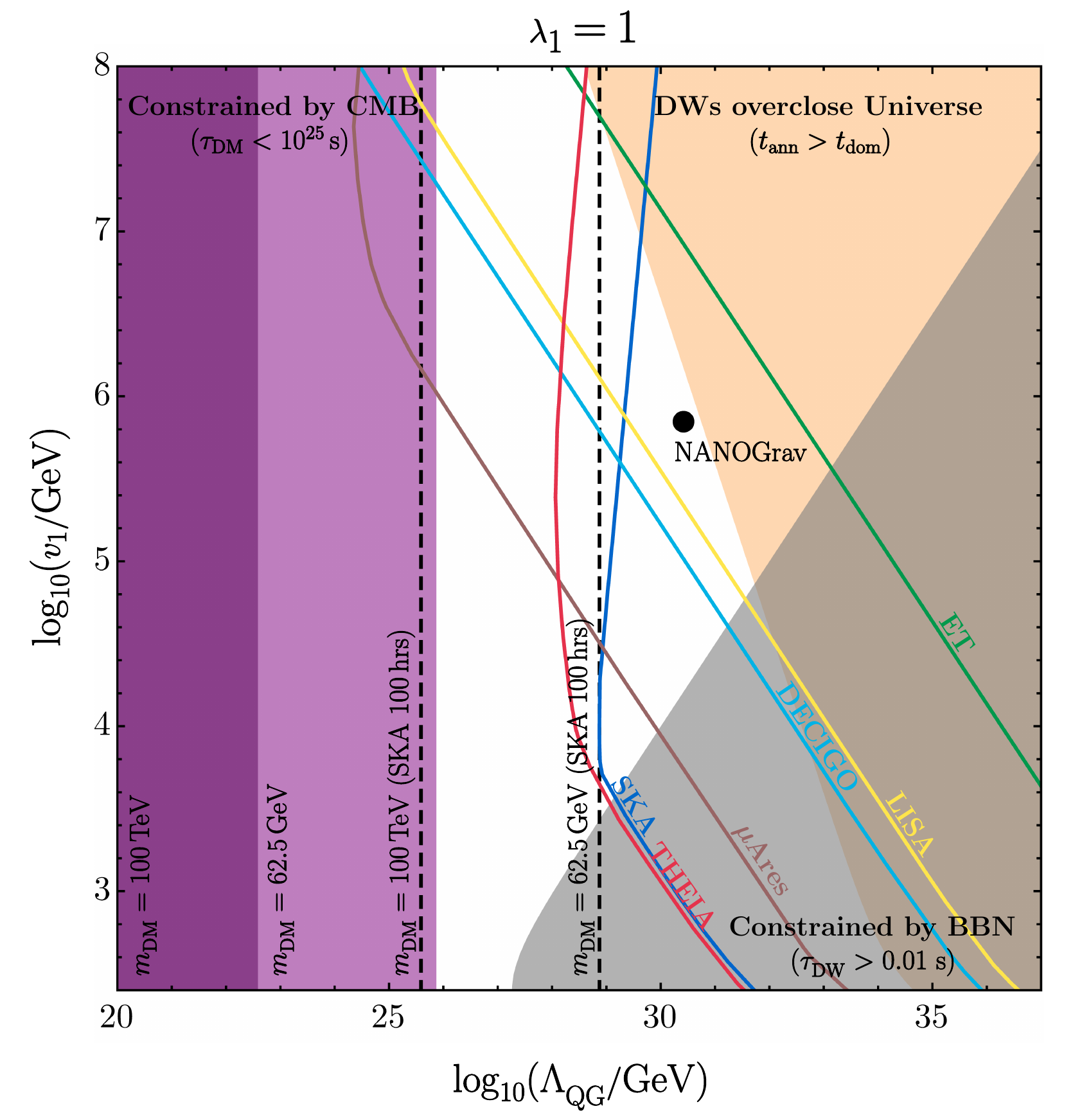}
  \includegraphics[width=0.32\linewidth]{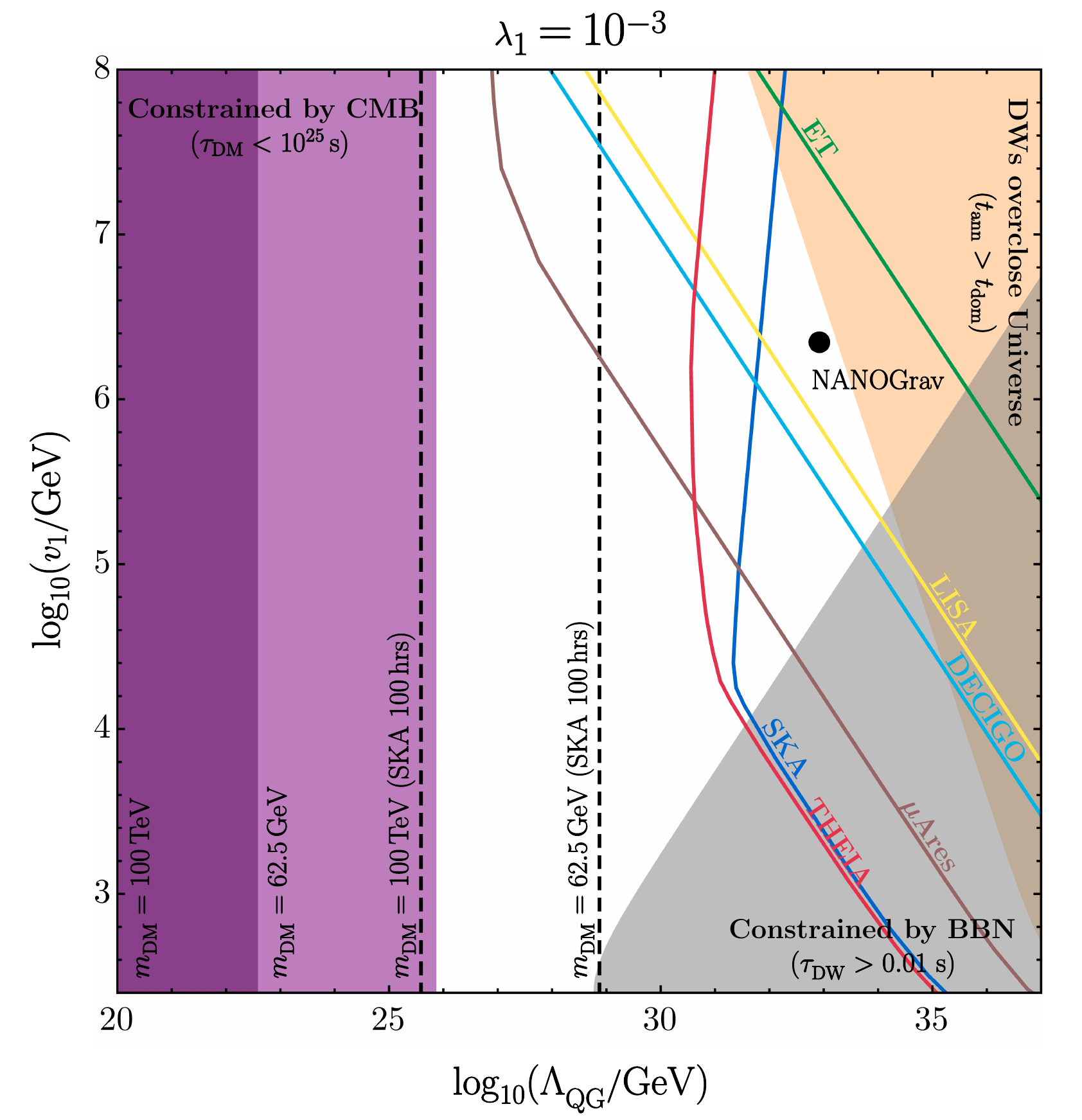}
  \includegraphics[width=0.32\linewidth]{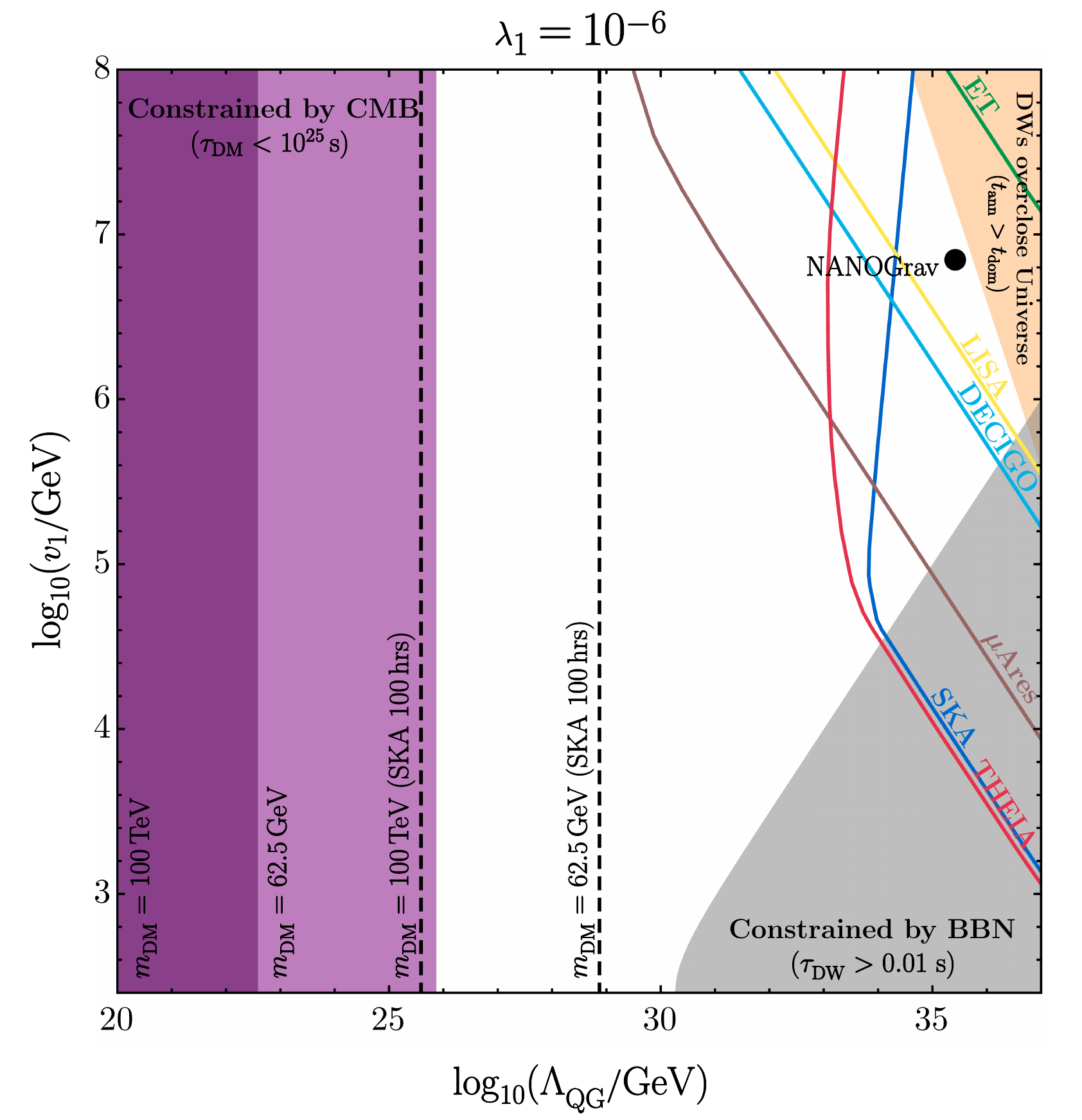}
  \caption{Combined constraints on $\Lambda_{\rm QG}$ and $v_1$ from indirect DM detection and GWs observations with varying $\lambda_1$. The darker and lighter purple-shaded regions denote the excluded regions by CMB observations, where we take $m_{\rm DM}$ to be 62.5~${\rm MeV}$ (Higgs resonance) and 100~${\rm TeV}$ (unitarity bound), respectively. The black dashed lines label the testing capabilities of the upcoming SKA telescope. The black points denote a benchmark with $f_p = 4.07 \times 10^{-8}~{\rm Hz}$ and $\Omega_p h^2 = 1.76\times10^{-7}$, which is consistent with NANOGrav 15-year results. The red, blue, brown, cyan, yellow and green curves present the testing capabilities of THEIA, SKA, $\mu$Ares, DECIGO, LISA and ET with ${\rm SNR}=10$. The gray-shaded regions are excluded by the requirement $\tau_{\rm DW} \lesssim 0.01~{\rm s}$. The orange-shaded regions correspond to the scenario where DWs may overclose the universe at an early epoch.}
  \label{fig:summary}
\end{figure*}

\vspace{0.5 em} 
\section{DWs annihilation and GWs}The spontaneous breaking of the $\ZDW$-symmetry could lead to the formation of 2D topological defects called DWs, filling our universe with patches in different degenerate vacua~\cite{Kibble:1976sj}. DWs are problematic as their energy density attenuates more slowly than that of radiation and matter, and would eventually become dominant. Hence DWs can alter the evolution of our universe in a way inconsistent with current CMB observations~\cite{Zeldovich:1974uw}. However, dimension-five operators associated with $S^{}_1$ in eq.~\eqref{eq:bias} can generate the energy bias term
\begin{equation}
    V^{}_{\rm bias} \simeq \frac{1}{\Lambda^{}_{\rm QG}}\left( v^5_1 + \frac{v^3_1 v^2_{h}}{2} + \frac{v^{}_1 v^4_{h}}{4} \right) \; ,
    \label{eq:model-bias}
\end{equation}
which softly breaks $\ZDW$-symmetry and also the vacua degeneracy. The population of the ``true vacuum'' (with lower energy) $p_{-}$ then should be greater than that of the ``false vacuum'' (with higher energy) $p_{+}$, i.e. $p_{+}/p_{-} \simeq {\rm exp}[-4V_{\rm bias}/(\lambda_1 v^4_1)]$~\cite{Gelmini:1988sf}. The different population between the two vacua induces a volume pressure force $p_V \sim V_{\rm bias}$ acting on the walls, forcing the false vacuum region to shrink. When $p_V$ is greater than the tension force $p_T$ of the walls, the DWs start to collapse and annihilate, triggering a characteristic SGWB signal. To achieve an observable GW signal, an enormous hierarchy between $V$ and $V_{\rm bias}$ is usually assumed by fiat. We emphasize that such a hierarchy is the natural consequence of the QG effects. It should also be mentioned that $v_1 \gg v_{h}$ is satisfied in most of the parameter space in our minimalistic model, hence a strong hierarchy among these operators should exist, rendering $v^5_1/\Lambda_{\rm QG}$ to be the dominant one.

The annihilation of DWs was investigated in, e.g. refs.~\cite{Vilenkin:1981zs, Gelmini:1988sf, Larsson:1996sp, Hiramatsu:2013qaa, Hiramatsu:2012sc, Saikawa:2017hiv}. Assuming the DWs annihilate in the radiation-dominated era, the peak frequency $f_p^{}$ and peak energy density  $\Omega^{}_{p}h^2_{}$ of GWs can be calculated at the annihilation time $t_{\rm ann}$ when $p_V \sim p_T$~\cite{Saikawa:2017hiv, Chen:2020wvu}
\begin{equation}
\begin{split}
f_p^{} & \simeq 3.75 \times 10^{-9}_{}~{\rm Hz} ~C^{-1/2}_{\rm ann} \mathcal{A}^{-1/2}_{} \widehat{\sigma}^{-1/2}_{} \widehat{V}^{1/2}_{\rm bias} \; , \\ 
\Omega^{}_{p} h^2_{} & \simeq 5.3 \times 10^{-20}_{}~ \widetilde{\epsilon} \mathcal{A}^4_{} C^2_{\rm ann} \widehat{\sigma}^{4}_{} \widehat{V}^{-2}_{\rm bias}  \; ,
\end{split}
\label{eq:peak-a}
\end{equation}
where $\widehat{\sigma}: = \sigma/{\rm TeV}^3$ with $\sigma: = \sqrt{8\lambda^{}_1 /9}\, v_1^3$ being the surface tension and $\widehat{V}_{\rm bias}: = V_{\rm bias}/{\rm MeV^4}$. The area parameter $\mathcal{A}$ and the dimensionless constant $C_{\rm ann}$ are respectively chosen as 0.8 and 2, and the efficiency parameter $\widetilde{\epsilon} \simeq 0.7$ can be regarded as a constant in the scaling regime~\cite{Hiramatsu:2012sc}. The degrees of freedom for energy and entropy density $g_{*s}(T_{\rm ann}) \simeq g_{*}(T_{\rm ann}) \simeq 10$ is taken into consideration for annihilation temperature $T_{\rm ann} =(1 \cdots 100)~{\rm MeV}$. Eq.~\eqref{eq:peak-a} indicates that $\Omega _p h^2 \propto v_1^2$. So even though string theory could result in many moduli, the GW signal will be dominated by the scalar with the largest vev. Therefore, our analysis captures the phenomenology of a large class of string-inspired models.

Two further remarks on DWs are as follows. First, if DWs annihilate into SM particles, they could dramatically reshape the era of BBN. In order to avoid this problem, we should require the lifetime $\tau_{\rm DW}$ to be shorter than $ t_{\rm ann} \lesssim 0.01\,{\rm s}$, which results in a lower bound on $V^{}_{\rm bias}$, namely, $V^{1/4}_{\rm bias} \gtrsim 5.07\times10^{-4}_{}~{\rm GeV}~C^{1/4}_{\rm ann}\mathcal{A}^{1/4}_{}\widehat{\sigma}^{1/4}_{}$. Converting into the constraint on $f^{}_p$, we arrive at $f^{}_{p} \gtrsim 0.964\times 10^{-9}~{\rm Hz}$, consistent with the latest SGWB signal detected by NANOGrav~\cite{NANOGrav:2023gor, NANOGrav:2023hvm}. Second, the domination of DWs in the early universe is generally allowed as long as they annihilate before BBN, but we have little knowledge about the dynamics of DWs in the DW-dominated universe. Then it would be useful to identify in which case $t_{\rm ann}$ is shorter than the time when DWs overclose the universe $t_{\rm dom}$. This would lead to another condition $V^{1/4}_{\rm bias} \gtrsim 2.18 \times 10^{-5}~{\rm GeV}~C^{1/4}_{\rm ann} \mathcal{A}^{1/2}_{} \widehat{\sigma}^{1/2}_{}$~\cite{Saikawa:2017hiv}.


\vspace{0.5 em}
\section{Combined constraints on the QG scale}\cref{fig:summary} summarizes our results in the $\{v_1,\Lambda_{\rm QG}\}$ parameter space. For DM, we consider two values of $m_{\rm DM}$, namely, $m_\text{DM}=62.5~\text{GeV}$ (Higgs resonance) and $100~\text{TeV}$ (unitarity bound)~\footnote{A detailed analysis of dark matter phenomenology in the two-multiplet extension of the scalar sector can be found in refs.~\cite{Basak:2021tnj,DuttaBanik:2020jrj}.}. Constraints from CMB observations are denoted by dark and light purple-shaded regions, respectively. One can find that CMB observations set a stringent lower bound on $\Lambda_{\rm QG}$. In particular, $\Lambda_{\rm QG} \gtrsim 10^{25.8}~{\rm GeV}$ for $m_\text{DM}=62.5~\text{GeV}$. The upcoming SKA radio telescope also provides a detection prospect of decaying DM, and a recent study~\cite{Dutta:2022wuc} suggests that $\Gamma_\text{DM}\gtrsim 10^{-30}~{\rm s}^{-1}$ is detectable by SKA, indicating $\Lambda_{\rm QG}$ up to $10^{29}~{\rm GeV}$ may be tested in the future. We use black dashed lines to illustrate this. 

\begin{figure}[t!]
  \centering
  \includegraphics[width=1\linewidth]{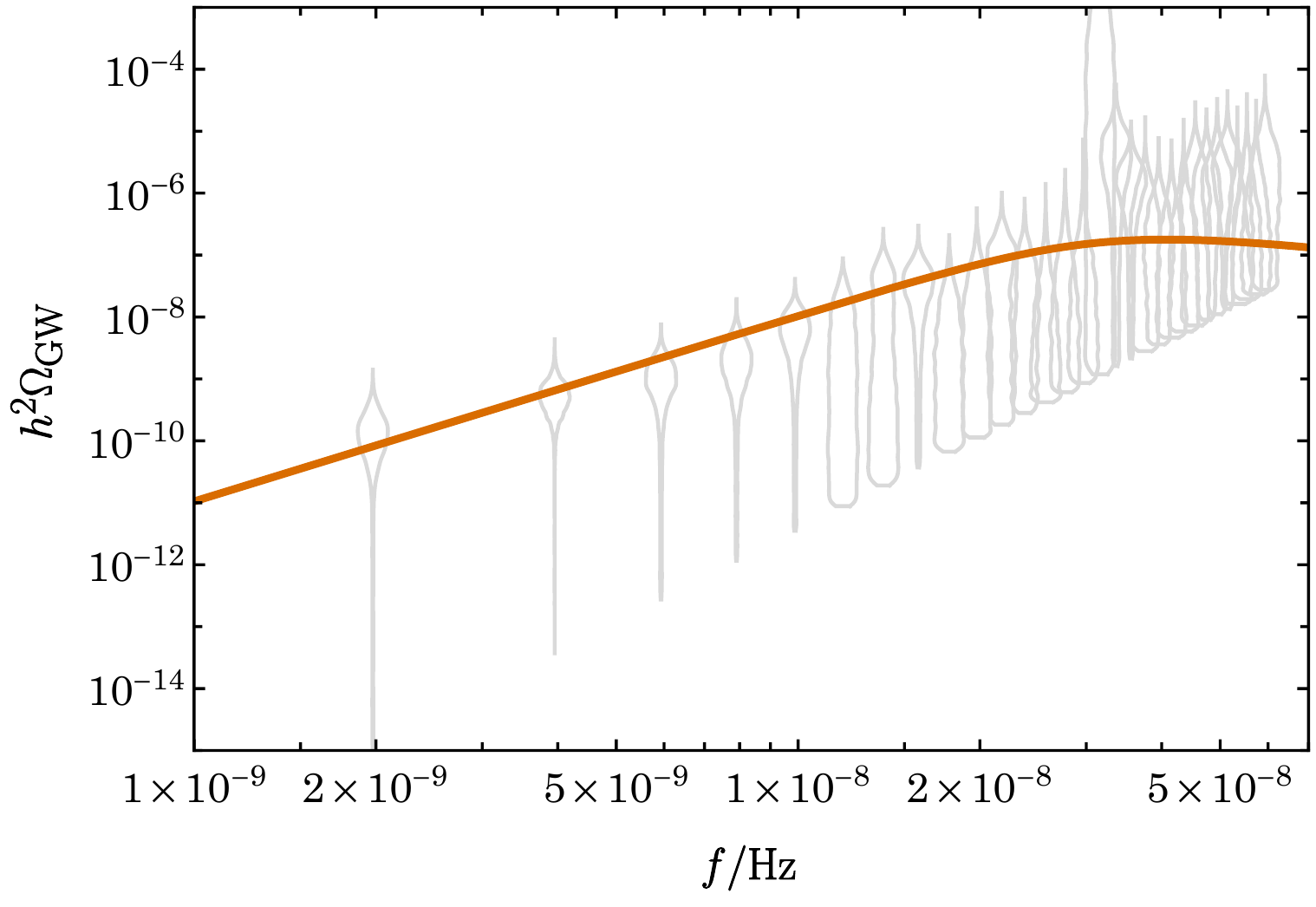}  \caption{Our benchmark GW spectrum is denoted as the orange curve, while gray violins show the NANOGrav 15-year data~\cite{NANOGrav:2023gor, NANOGrav:2023hvm}.}
  \label{fig:fit-DW2}
\end{figure}

As for SGWB from DWs, we choose $\lambda_1=1$, $10^{-3}$ and  $10^{-6}$ for demonstration. The black points in~\cref{fig:summary} represents 
a benchmark with $f_p = 4.07 \times 10^{-8}~{\rm Hz}$ and $\Omega_p h^2 = 1.76\times10^{-7}$, which is consistent with NANOGrav 15-year results~\cite{NANOGrav:2023gor, NANOGrav:2023hvm}. Taking $\lambda_1 = 10^{-3}_{}$ for instance, it corresponds to $\Lambda_{\rm QG} = 8.68 \times 10^{32}~{\rm GeV}$ and $v_1 = 2.21 \times 10^6~{\rm GeV}$. In order to depict the GW spectrum, we adopt the following parametrization for a broken power-law spectrum~\cite{Caprini:2019egz, NANOGrav:2023hvm}
\begin{eqnarray}
h^2_{} \Omega^{}_{\rm GW} = h^2_{} \Omega^{}_p \frac{(a+b)^c}{\left(b x^{-a / c}+a x^{b / c}\right)^c} \ ,
\label{eq:spec-par}
\end{eqnarray}
where $x := f/f_p$, and $a$, $b$ and $c$ are real and positive parameters. Here the low-frequency slope $a = 3$ can be fixed by causality, while numerical simulations suggest $b \simeq c \simeq 1$~\cite{Hiramatsu:2013qaa}. The corresponding GW spectrum is shown in \cref{fig:fit-DW2} using an orange curve, with gray violins denoting NANOGrav's results for comparison. Moreover, the gray-shaded regions in \cref{fig:summary} are excluded by the requirement that DWs should annihilate  before BBN ($\tau_\text{DW}\lesssim 0.01~\text{s}$), which sets a restriction on large $\Lambda_{\rm QG}$. This constraint can be stronger as $\lambda_1$ goes larger. The orange-shaded regions correspond to the scenario where DWs may overclose the universe at an early epoch ($t_\text{ann}>t_\text{dom}$). We can observe that our benchmark points are very close to the boundaries of these regions. In addition, we also investigate the capabilities of other GW detectors for constraining QG effects. We calculate the signal-to-noise ratio (SNR)~\cite{Maggiore:1999vm,Allen:1997ad} for ET~\cite{Punturo:2010zz}, LISA~\cite{LISA:2017pwj}, DECIGO~\cite{Kawamura:2020pcg}, $\mu$Ares~\cite{Sesana:2019vho}, SKA~\cite{Janssen:2014dka} and THEIA~\cite{Garcia-Bellido:2021zgu} detectors by imputing the spectrum given in eq.~\eqref{eq:spec-par}, and plot their individual sensitivity curves with $\rm{SNR}=10$ by~\cref{fig:summary}. The regions bounded by these curves refer to the peak frequencies and energy densities with which the GW spectra can be tested. Broader areas are enclosed in the $\{v_1,\Lambda_{\rm QG}\}$ parameter space for larger $\lambda_1$. Most of these curves contain the NANOGrav benchmark point, therefore it is promising that these GW detectors can give us combined constraints on $\Lambda_{\rm QG}$ and $v_1$ in a multi-frequency range. 

\vspace{0.5 em} 
\section{Conclusion}In this work we have argued that a large class of string-inspired models has phenomenology that can plausibly lead us to measure the effective scale of quantum gravity. We have considered the low energy consequences of the swampland conjecture that global symmetries are broken by quantum gravity---that dark matter and domain walls can both become metastable as a result. The scale of quantum gravity effects that can be measured corresponds to a plausible range of values for the wormhole action, ${\cal S}\sim (4 \cdots 38)$. If the phenomenology mentioned in this paper is seen, it provides evidence for the paradigm of non-perturbative quantum-gravity instantons breaking global symmetries. A tantalizing possibility is that recent observations of a gravitational wave spectrum by pulsar timing arrays were produced by primordial metastable domain walls, and perhaps the gravitational wave spectrum is our first empirical information about quantum gravity.

\vspace{0.5 em} 
\noindent{\it Acknowledgements}.---We would like to thank Volodymyr Takhistov for his collaboration in the early stages of this project. SFK and GW acknowledge the STFC Consolidated Grant ST/L000296/1 and SFK also acknowledges the European Union's Horizon 2020 Research and Innovation programme under Marie Sklodowska-Curie grant agreement HIDDeN European ITN project (H2020-MSCA-ITN-2019//860881-HIDDeN). 
RR acknowledges financial support from the STFC Consolidated Grant ST/T000775/1. 
XW acknowledges the Royal Society as the funding source of the Newton International Fellowship.
MY was supported in part by the JSPS Grant-in-Aid for Scientific Research (No.\ 19H00689, 19K03820, 20H05860, 23H01168), and by JST, Japan (PRESTO Grant No. JPMJPR225A, Moonshot R\&D Grant No.\ JPMJMS2061).

\bibliographystyle{apsrev4-1}
\bibliography{refs}

\onecolumngrid
\appendix







\end{document}